\documentclass{article}
\usepackage[preprint]{neurips_2026}

\usepackage[utf8]{inputenc}
\usepackage[T1]{fontenc}
\usepackage{hyperref}
\usepackage{url}
\usepackage{booktabs}
\usepackage{multirow}
\usepackage{amsfonts}
\usepackage{nicefrac}
\usepackage{microtype}
\usepackage{xcolor}
\usepackage{colortbl}
\usepackage{array}
\usepackage{hhline}
\usepackage{tikz}
\usetikzlibrary{tikzmark,calc}

\usepackage{graphicx}
\usepackage{amsmath}
\usepackage{amssymb}
\usepackage{mathtools}
\usepackage{amsthm}
\usepackage{verbatim}
 
\theoremstyle{definition}
\newtheorem{example}{Example}
\newtheorem{remark}{Remark}

\newcommand{\dd}{\mathrm{d}}

\title{A renormalization-group inspired lattice-based framework
 for piecewise generalized linear models}

\author{%
  Joshua C.\ Chang \\
  NIH Clinical Center, Rehabilitation Medicine \\
  \texttt{josh.chang@nih.gov}\\ \\
  Mederrata Research Inc \\ 
    \texttt{josh@mederrata.org} \\ \\
  Sound Prediction Inc \\  
    \texttt{josh@soundprediction.com}
}

\begin{document}

\maketitle

\begin{abstract}
We formally introduce a class of models inspired by renormalization group (RG) theory, built on additive hierarchical expansions analogous to those appearing in functional ANOVA and mixed-effects models.
Like ReLU convolutional neural networks, they are almost everywhere locally linear; unlike ReLU networks, their partition structure is explicit, interpretable, and easy to modify or constrain.
In these models, one defines a multidimensional lattice partition of the input space and uses it to scaffold variations in regression parameters.
Each dimension of the lattice corresponds to an attribute by which the statistics of the problem may vary.
The parameters are themselves expressed in the form of an expansion, where each term captures variations relative to a lower (coarser) interaction scale.
These models admit multiple equivalent interpretations: as piecewise GLMs, as hierarchical mixed-effects regressions, or as regression trees with structured parameter sharing.
Since RG motivates the design of these models, we use techniques from statistical physics -- specifically replica analysis -- to study their generalization properties. Specifically, we analyze the behavior of the Watanabe-Akaike Information Criterion (WAIC) as a proxy for generalization loss.
This analysis yields two practical results: (i) guidance on the lattice design as a function of dataset size and predictor dimensionality; and (ii) a principled scaling law for the regularization prior when adding higher-order terms to the expansion so that one can increase model complexity without an expected increase in generalization loss.
We evaluate the methodology on public datasets and find performance competitive against both blackbox methods and other intrinsically interpretable approaches.

\end{abstract}

\section{Introduction}

\begin{figure}[h]
    \centering
    \includegraphics[width=0.76\linewidth]{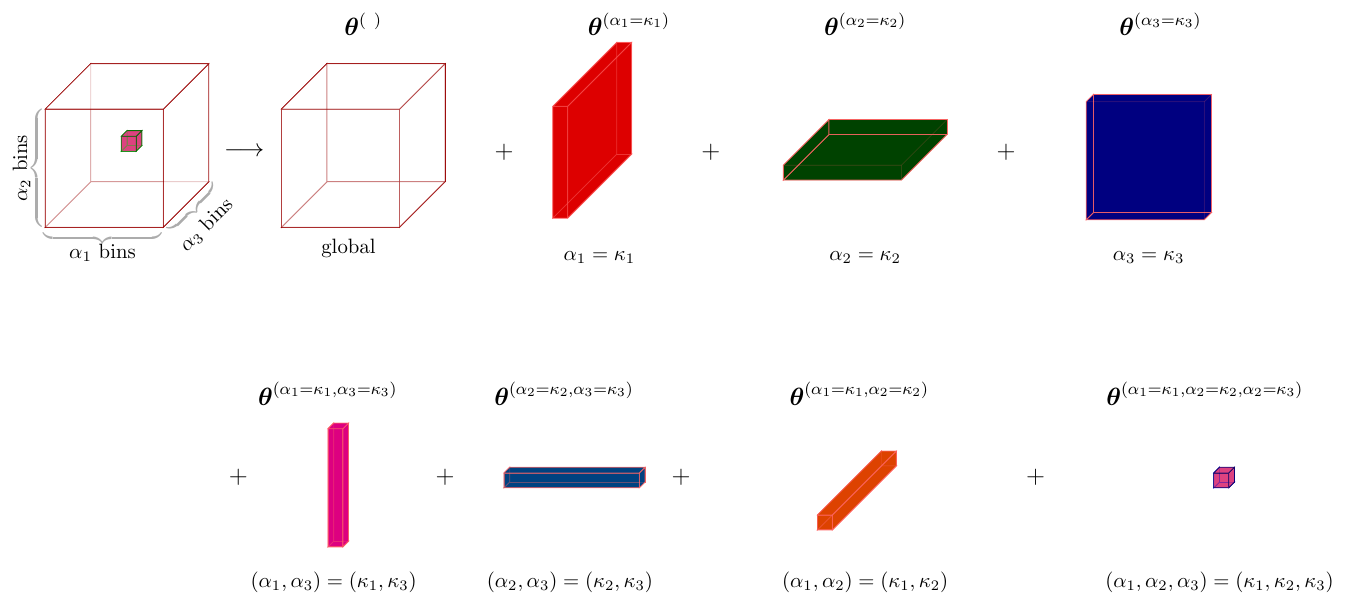}
    \caption{\textbf{Decomposing a parameter} by global, mesoscopic, and local interactions}
    \label{fig:decomposition}
\end{figure}

In prior applied work, some authors have used a particular form of additive hierarchical expansion, termed \emph{Bayesianquilts}, to represent changes in model parameters as a sum of contributions across different interaction scales~\citep{changInterpretableNotJust2024,xiaInterpretableNotJust2023,changFunctionalImprovementBetter2025,hoEffectivenessTicketWork2025}.
Here we formally introduce this model class and analyze its generalization properties.
The motivation behind these models is in maintaining strict intrinsic interpretability while allowing effects to vary (nonlinearity).
Under the idea that nearby (in some sense) observations behave similarly, this modeling technique considers multiple ways in which to partition data points.
Together these partitionings define a lattice (Fig.~\ref{fig:decomposition}), which when sliced in different ways define length scales in the dataset.
These models decompose parameters across scales in a manner analogous to functional ANOVA~\citep{hoeffdingClassStatisticsAsymptotically1948,SobolSensitivityEstimation}, where effects are partitioned into global, mesoscopic, and local interaction components.
The resulting structure admits multiple equivalent interpretations: as a piecewise generalized linear model (GLM) over an explicit lattice partition, as a hierarchical mixed-effects model with nested random effects, or as a regression tree with structured parameter sharing across cells.

A regression model is \emph{piecewise GLM} if the domain partitions into regions, each governed by a local GLM with cell-specific linear predictor $\eta(\mathbf{x}) = \mathbf{w}(\mathbf{x}) \cdot \mathbf{x} + b(\mathbf{x})$ for input $\mathbf{x} \in \Omega \subseteq \mathbb{R}^p$, where $\mathbf{w}(\mathbf{x})$ and $b(\mathbf{x})$ are the cell-specific weight vector and intercept, and $g$ is a link function relating $\eta$ to the expected response.
ReLU convolutional neural networks belong to a related piecewise linear family, with partitions defined implicitly by activation patterns~\citep{sudjiantoUnwrappingBlackBox2020,saitoExtractingRegressionRules2002}.
Decision trees and their ensembles (Random Forest, XGBoost) similarly partition the domain, though adaptively via recursive splitting.
The models we introduce share this piecewise structure but differ in two key respects: the partition is explicit and predetermined on interpretable features, and parameters share structure across cells through a hierarchical decomposition.
These models are intrinsically interpretable~\citep{rudinStopExplainingBlack2019}.
Post-hoc explainability methods like LIME~\citep{ribeiroWhyShouldTrust2016} and SHAP~\citep{lundbergUnifiedApproachInterpreting2017} can approximate black-box predictions locally, but they fail to capture mesoscopic model structure and can be misleading~\citep{rudinWhyBlackBox2022,aasExplainingIndividualPredictions2021,adebayoPostHocExplanations2022,alvarez-melisRobustnessInterpretabilityMethods2018,bordtPostHocExplanationsFail2022}; the insurmountable limitations of posthoc explainability methods motivate the need for intrinsically interpretable models.

Concretely, suppose each input maps to a cell indexed by $\boldsymbol\kappa \in \mathbb{N}^d$ within a lattice, and let $\boldsymbol\alpha$ index locations within sublattices at each interaction level.
We decompose the cell-specific parameter as:
\begin{align}
    \boldsymbol{\theta}^{\boldsymbol\kappa}
    &= \boldsymbol{\theta}^{(\ )} + \sum_{i} \boldsymbol{\theta}^{(\alpha_i=\kappa_i)} + \sum_{i<j} \boldsymbol{\theta}^{(\alpha_i=\kappa_i,\alpha_j=\kappa_j)} + \dots,
    \label{eq:decomposition}
\end{align}
where each term $\boldsymbol{\theta}^{(\boldsymbol\alpha)}$ captures variation at a particular scale -- global intercept, main effects along each lattice dimension, pairwise interactions, and so on (Fig.~\ref{fig:decomposition}).
This structure induces partial pooling: finer-scale terms are regularized more strongly, so data-sparse cells borrow strength from coarser groupings.
The lattice can be chosen by domain experts to encode domain knowledge and can consist of categorical variables, binned continuous variables, or binned latent representations of the input features~\citep{changSparseEncodingMoreinterpretable2021}. For example, a clinical model might partition patients by age, sex, and discretizations of latent representations for medical utilization.

Our central contribution is using techniques from statistical physics to study the generalization properties of this model class.
We apply replica analysis -- a trick for computing a complicated expectation as a more tractable limit -- to compute the expected generalization loss and its dependence on model complexity.
This analysis yields two practical results (writing $N$ for sample size and $p$ for the number of regression parameters):

\textbf{1) Bin size:} When discretizing $d_{\text{cont}}$ continuous covariates into $L$ bins each, the local parameter-to-sample ratio $\gamma_{\text{local}} = p L^{d_{\text{cont}}} / N$ must remain below unity for the replica-symmetric approximation to hold, constraining $L < (N/p)^{1/d_{\text{cont}}}$.
\textbf{2) Generalization-preserving regularization:} In the Gaussian case we derive a scaling law for the prior standard deviation $\tau^{(\boldsymbol\alpha)} \leq \sigma / \sqrt{2p \cdot N^{(\boldsymbol\alpha)}}$, where $\sigma$ is the noise scale and $N^{(\boldsymbol\alpha)}$ is the local sample size, such that adding higher-order terms does not increase expected generalization loss, even when the true effect is zero. we then extend this criterion to GLMs in general.

\subsection{Related work}

\textbf{Piecewise and additive models.}
Locally weighted regression~\citep{clevelandLocallyWeightedRegression1988} and MARS~\citep{friedmanMultivariateAdaptiveRegression1991} introduced piecewise structures for flexible nonparametric modeling.
Generalized additive models (GAMs)~\citep{GeneralizedAdditiveModels} provide hierarchical penalization of main effects versus interactions through continuous basis expansions.
Explainable Boosting Machines (EBM)~\citep{noriInterpretMLUnifiedFramework2019,caruanaIntelligibleModelsHealthCare2015,louAccurateIntelligibleModels2013} use an additive decomposition with main effects and pairwise interactions learned via gradient boosting.
Semi-structured deep piecewise models~\citep{kopperSemiStructuredDeepPiecewise2021} combine neural networks with piecewise components.
Our approach differs in using an explicit lattice partition where each cell corresponds to a named subgroup, with hierarchical parameter sharing that provides theoretical guarantees on regularization scaling.

\textbf{Statistical physics of learning.}
Replica methods were applied to neural networks by~\citet{gardnerSpaceInteractionsNeural1988} and to generalization by~\citet{seungStatisticalMechanicsLearning1992}.
The replica calculation of effective degrees of freedom for ridge regression under random design is classical~\citep{kroghGeneralizationLinearPerceptron1992}.
A large body of work connects machine learning to RG through exact mappings~\citep{mehtaExactMappingVariational2014,linWhyDoesDeep2017}, neural network architectures~\citep{benyDeepLearningRenormalization2013,liNeuralNetworkRenormalization2018}, and information-theoretic coarse-graining~\citep{shwartz-zivOpeningBlackBox2017,koch-januszMutualInformationNeural2018}.
Our contribution is not the replica calculation itself, but its application to hierarchical decompositions: we derive how regularization should scale across interaction orders (Eq.~\ref{eq:scaling-law}) so that adding finer-scale terms is not expected to hurt generalization, and we identify the optimal truncation order as a critical point in an RG flow (Eq.~\ref{eq:critical-order}).

\section{Preliminaries}

\subsection{Notation}

Let $N$ denote the total sample size and $n \in \{1, \ldots, N\}$ index observations.
Let $p$ denote the number of regression coefficients (feature dimension) and $d$ the number of lattice dimensions (grouping factors).
Each lattice dimension has $L$ levels, yielding $L^d$ cells at full resolution (in practice we allow $L$ to vary between dimensions but keep it uniform here to keep the derivations simple).
We write $K$ for the truncation order in the hierarchical expansion and $k \in \{0, 1, \ldots, K\}$ for interaction order.
For a component indexed by $\boldsymbol\alpha$, let $\Omega^{(\boldsymbol\alpha)}$ denote the set of observations in that subset, $N^{(\boldsymbol\alpha)} = |\Omega^{(\boldsymbol\alpha)}|$ the local sample size, and $\pi^{(\boldsymbol\alpha)} = N^{(\boldsymbol\alpha)}/N$ the fraction of data.

The prior standard deviation for component $\boldsymbol\alpha$ is $\tau^{(\boldsymbol\alpha)}$; the noise standard deviation (Gaussian case) is $\sigma$.
In the RG flow analysis, $\rho \in (0,1)$ denotes the decay rate of effect sizes across scales: $(\theta^{(k)}_*)^2 \sim \rho^k$.
We write $\gamma = p/N$ for the parameter-to-sample ratio; locally, $\gamma^{(\boldsymbol\alpha)} = p/N^{(\boldsymbol\alpha)}$.

\subsection{Generalization error and WAIC}
\label{sec:waic-prelim}

For singular models like neural networks, where the mapping from parameters to distributions is not one-to-one, the Watanabe-Akaike Information Criterion (WAIC) provides an asymptotic approximation of the leave-one-out cross-validation error~\citep{watanabeWidelyApplicableBayesian2013,vehtariPracticalBayesianModel2017}:
\begin{equation}
    S^{\text{WAIC}} = -\sum_{n=1}^N \log \mathbb{E}_{\boldsymbol{\theta}|y,X}[f(y_n|x_n, \boldsymbol{\theta})] + \sum_{n=1}^N \text{Var}_{\boldsymbol{\theta}|y,X}[\log f(y_n|x_n, \boldsymbol{\theta})].
    \label{eq:waic-prelim}
\end{equation}
The first term is the "functional training error," and the second is a complexity penalty (analogous to the penalty term of the AIC).
This "action" $S^{\text{WAIC}}$ approximates model stability across scales.

To derive scaling laws for regularization, we borrow techniques from statistical physics.
The log-posterior corresponds to a Hamiltonian (energy function) $H(\boldsymbol{\theta}) = -\log f(y|x,\boldsymbol{\theta}) - \log \pi(\boldsymbol{\theta})$, with marginal likelihood as partition function (normalizing constant) $Z = \int \exp(-H(\boldsymbol{\theta})) \dd\boldsymbol{\theta}$.
Parameters $\boldsymbol{\theta}$ are thermal variables (quantities that fluctuate); data $(y, X)$ are quenched disorder (fixed randomness that defines the problem instance -- i.e., the training set is drawn once and held fixed).
Typical-case analysis requires averaging $\mathbb{E}_{\text{data}}[\log Z]$ rather than $\log \mathbb{E}_{\text{data}}[Z]$ -- a technically challenging quenched average addressed via replica methods (Appendix~\ref{app:replica}).
The RG, originally developed to study phase transitions~\citep{wilsonRenormalizationGroupCritical1983}, has been applied to deep learning as hierarchical coarse-graining~\citep{mehtaExactMappingVariational2014}.

\section{Methods}
Consider the Gaussian regression case, where 
one models each term in the hierarchical decomposition (Eq.~\ref{eq:decomposition}) using the prior $\boldsymbol{\theta}^{(\boldsymbol\alpha)} \sim \mathcal{N}(\mathbf{0}, (\tau^{(\boldsymbol\alpha)})^2 \mathbf{I})$.
The prior standard deviation $\tau^{(\boldsymbol\alpha)}$ determines the effective degrees of freedom $\text{df}_{\text{eff}}^{(\boldsymbol\alpha)}$ contributed by each component.
Our goal is to choose regularization at each scale such that adding terms does not hurt generalization in expectation when the true effect is zero.
The extension from the Gaussian case to GLMs follows via Fisher information (Section~\ref{sec:glm-extension}).

Consider adding a parameter $\theta^{(\boldsymbol\alpha)}$ observed through $N^{(\boldsymbol\alpha)}$ observations with sample mean $\bar{y}^{(\boldsymbol\alpha)}$.
Under the prior $\theta^{(\boldsymbol\alpha)} \sim \mathcal{N}(0, \tau^2)$, the posterior mean is shrunk by factor
\begin{equation}
    s = \frac{N^{(\boldsymbol\alpha)} \tau^2}{N^{(\boldsymbol\alpha)} \tau^2 + \sigma^2},
    \label{eq:shrinkage}
\end{equation}
toward zero: $\hat\theta = s \cdot \bar{y}^{(\boldsymbol\alpha)}$.
The WAIC complexity penalty -- the effective degrees of freedom -- equals this shrinkage factor $s$ (see Appendix~\ref{app:generalization-preserving}).
When the true effect is zero, adding the parameter contributes expected complexity cost $s$ without improving fit.
For a component with $p$ regression coefficients sharing the same prior variance, total $\text{df}_{\text{eff}} \approx p \cdot s$.
Bounding this to $1/2$ requires $s \leq 1/(2p)$; solving Eq.~\ref{eq:shrinkage} yields:
\begin{equation}
    \tau^{(\boldsymbol\alpha)}
    \leq \frac{\sigma}{\sqrt{2p \cdot N^{(\boldsymbol\alpha)}}}
    = \frac{\sigma}{\sqrt{2p \cdot N \cdot \pi^{(\boldsymbol\alpha)}}},
    \label{eq:scaling-law}
\end{equation}
where $N^{(\boldsymbol\alpha)}$ is the local sample size and $\pi^{(\boldsymbol\alpha)} = N^{(\boldsymbol\alpha)}/N$ is the fraction of data in the component.
This bound is an approximation based on expected behavior; in practice it serves as a principled default that can be tuned via cross-validation.
Adding a component improves generalization only when the signal exceeds the regularization-induced complexity cost; for $G$ groups with local sample sizes $N_g$, improvement requires $\sum_g N_g \|\boldsymbol{\beta}_{g,*}\|^2/(2\sigma^2) > Gp/2$.

For a $d$-dimensional lattice with $L$ levels per factor, the order-$k$ standard deviation scales as $\tau^{(k)} \propto \sqrt{L^k/N}$ (from the scaling law Eq.~\ref{eq:scaling-law} with $N^{(k)} = N/L^k$).
The prior variance increases geometrically with order: finer-scale terms are permitted larger fluctuations because they are constrained by fewer observations.
This prior scaling is distinct from the true effect decay rate $\rho$: the prior permits $(\tau^{(k)})^2 \propto L^k$ while true effects may decay as $(\theta^{(k)}_*)^2 \propto \rho^k$ for $\rho < L$.

\subsection{Renormalization group flow and optimal truncation}
\label{sec:rg-flow}

The sequence of truncated models defines a discrete RG flow where the truncation order $K$ plays the role of an inverse length scale.
In physics, the RG describes how effective theories change under coarse-graining -- averaging over fine-scale fluctuations to obtain coarser descriptions.
Here, increasing $K$ corresponds to resolving finer interaction scales, analogous to examining a system at shorter length scales.

We define the generalization gap $\Delta S_K = S^{\text{WAIC}}_K - S^{\text{WAIC}}_{K-1}$, measuring the change in generalization loss when including order-$K$ terms.
Since lower $S$ means better generalization, $\Delta S_K < 0$ (negative gap) means adding order-$K$ interactions \emph{improves} out-of-sample prediction, while $\Delta S_K > 0$ (positive gap) means it \emph{hurts} generalization.
In physics, the analogous quantity is called the ``beta function,'' which describes how coupling constants change under scale transformations; negative beta functions indicate relevant (important) interactions, positive ones indicate irrelevant (noise-dominated) interactions.

Assuming effect sizes decay with scale as $(\theta^{(k)}_*)^2 \sim \rho^k$ for $\rho < 1$, there exists a critical order $K^*$ satisfying $\Delta S_{K^*} = 0$.
In physics, a fixed point is where the system's behavior becomes scale-invariant; here, $K^*$ is the truncation order where adding more interactions neither helps nor hurts generalization -- the optimal bias-variance tradeoff.
As derived in Appendix~\ref{app:critical}, this fixed point occurs at:
\begin{equation}
    K^* \approx \frac{\log(N/\sigma^2)}{\log(L/\rho)}.
    \label{eq:critical-order}
\end{equation}
The sharpness of this transition depends on the ratio $\rho/L$: when $\rho \ll L$, the transition from underfitting to overfitting is sharp; when $\rho \to L$, it is gradual and $K^*$ is less well-defined.
The generalization-preserving regularization ensures the flow is stable beyond $K^*$, suppressing spurious interactions without requiring explicit model selection.

\textbf{Warm-start initialization.}
When refining a model from a coarser to a finer lattice, the parameters from the coarser model provide an effective initialization.
For a lattice refinement $\mathcal{L} \to \mathcal{L}'$ where each cell in $\mathcal{L}$ splits into multiple cells in $\mathcal{L}'$, we initialize the finer parameters by replicating the coarser values:
$\boldsymbol{\theta}^{(\alpha')} \gets \boldsymbol{\theta}^{(\pi(\alpha'))}$
where $\pi: \mathcal{L}' \to \mathcal{L}$ maps each fine cell to its parent coarse cell.
This warm-start approach accelerates convergence and helps avoid local minima by starting from a solution that already captures coarse-scale structure.
The same principle applies when increasing truncation order: parameters from an order-$K$ model initialize the corresponding terms in an order-$(K+1)$ model, with new interaction terms initialized to zero.

\textbf{Adaptive discretization.}
Continuous covariates can be incorporated as lattice dimensions via discretization.
The bin count is determined adaptively using the contingency table: if the current lattice has minimum cell count $n_{\min}$, a new dimension with $L$ levels is feasible only if $n_{\min}/L \geq n_{\text{threshold}}$, where $n_{\text{threshold}}$ ensures sufficient observations per cell for stable estimation.
This criterion derives from the generalization-preserving condition: finer binning is permitted where data is abundant, coarser where sparse.

\begin{remark}[Bin count for continuous covariates]
\label{rem:bin-count}
Consider discretizing $d_{\text{cont}}$ continuous covariates into $L$ bins each, with $p$ regression coefficients per cell.
The local sample size scales as $N^{(\boldsymbol\alpha)} \approx N / L^{d_{\text{cont}}}$ for balanced partitions, yielding a local parameter-to-sample ratio
\begin{equation}
    \gamma_{\text{local}} = \frac{p}{N^{(\boldsymbol\alpha)}} = \frac{p \cdot L^{d_{\text{cont}}}}{N}.
    \label{eq:gamma-local}
\end{equation}
The replica analysis (Appendix~\ref{app:replica}) assumes $\gamma < 1$ for the replica-symmetric saddle point to be valid.
When $\gamma_{\text{local}} > 1$ in Eq.~\ref{eq:gamma-local}, each cell is overparameterized: the posterior becomes multimodal, replica symmetry may break, and our mean-field predictions for the generalization loss no longer apply.
Requiring $\gamma_{\text{local}} < 1$ in Eq.~\ref{eq:gamma-local} and solving for $L$ yields the constraint
\begin{equation}
    L < \left( \frac{N}{p} \right)^{1/d_{\text{cont}}}.
    \label{eq:bin-constraint}
\end{equation}
For a single discretized covariate ($d_{\text{cont}}=1$), Eq.~\ref{eq:bin-constraint} gives $L_{\max} \sim N/p$.
For multiple discretized covariates, Eq.~\ref{eq:bin-constraint} tightens rapidly: with $d_{\text{cont}}=3$ and $N/p = 1000$, the constraint yields $L_{\max} \approx 10$; with $d_{\text{cont}}=5$, $L_{\max} \approx 4$.
Note that $d_{\text{cont}}$ counts only continuous features being discretized; pre-existing categorical features contribute to the lattice structure but not to this constraint, as their cell counts are fixed by the data rather than chosen.
In practice, we recommend $L \leq (N/p)^{1/d_{\text{cont}}} / 2$ to ensure $\gamma_{\text{local}} \leq 1/2$, providing a margin of safety for the replica-symmetric approximation.
\end{remark}

\subsection{Extension to generalized linear models}
\label{sec:glm-extension}

The generalization-preserving regularization extends naturally to generalized linear models (GLMs) by replacing the noise variance $\sigma^2$ with the inverse Fisher information.
For a GLM with canonical link function $g(\cdot)$ and mean $\mu_n = g^{-1}(\mathbf{x}_n^\mathsf{T}\boldsymbol\beta)$, the log-likelihood is
$\ell(\boldsymbol\beta) = \sum_{n=1}^N [ y_n \eta_n - b(\eta_n) ] / a(\phi)$,
where $\eta_n = \mathbf{x}_n^\mathsf{T}\boldsymbol\beta$ is the linear predictor and $b(\cdot)$ is the cumulant function.

The Fisher information matrix is $\mathcal{I}(\boldsymbol\beta) = \mathbf{X}^\mathsf{T} \mathbf{W} \mathbf{X}$, where $\mathbf{W} = \text{diag}(w_1, \ldots, w_N)$ with weights $w_n = b''(\eta_n) / a(\phi)$.
For logistic regression, $w_n = \mu_n(1 - \mu_n)$; for Poisson regression, $w_n = \mu_n$.

The effective degrees of freedom for a component $\boldsymbol\beta^{(\boldsymbol\alpha)}$ becomes
\begin{equation}
    \text{df}_{\text{eff}}^{(\boldsymbol\alpha)} = \text{tr}\left[ \left( \mathbf{X}_{(\alpha)}^\mathsf{T} \mathbf{W}_{(\alpha)} \mathbf{X}_{(\alpha)} + \tau^{-2} \mathbf{I} \right)^{-1} \mathbf{X}_{(\alpha)}^\mathsf{T} \mathbf{W}_{(\alpha)} \mathbf{X}_{(\alpha)} \right],
    \label{eq:glm-df}
\end{equation}
where $\mathbf{X}_{(\alpha)}$ and $\mathbf{W}_{(\alpha)}$ are restricted to observations in subset $\Omega^{(\boldsymbol\alpha)}$.

Bounding Eq.~\ref{eq:glm-df} by $1/2$ and solving for $\tau$, the generalization-preserving scaling becomes
\begin{equation}
    \tau^{(\boldsymbol\alpha)} \leq \frac{1}{\sqrt{2p \cdot \bar{w}^{(\boldsymbol\alpha)} \cdot N^{(\boldsymbol\alpha)}}},
    \label{eq:glm-scaling}
\end{equation}
where $\bar{w}^{(\boldsymbol\alpha)} = (N^{(\boldsymbol\alpha)})^{-1} \sum_{n \in \Omega^{(\boldsymbol\alpha)}} w_n$ is the average weight in the subset.
For balanced binary classification with $\mu_n \approx 0.5$, $\bar{w} \approx 0.25$, yielding $\tau \leq \sqrt{2}/\sqrt{p \cdot N^{(\boldsymbol\alpha)}}$ -- twice the Gaussian case (since $1/\sqrt{2 \cdot 0.25} = \sqrt{2}$).
For rare events with prevalence $\pi_+ \ll 1$, $\bar{w} \approx \pi_+$, yielding $\tau \leq 1/\sqrt{2p \cdot \pi_+ \cdot N^{(\boldsymbol\alpha)}}$.

In practice, we evaluate $\bar{w}^{(\boldsymbol\alpha)}$ at the current parameter estimates $\hat{\boldsymbol\beta}$ and update the regularization adaptively during optimization.
This iteratively reweighted approach converges to a fixed point where the regularization is consistent with the fitted model (see Appendix~\ref{app:glm} for details).

\begin{example}[Hierarchical logistic regression]
\label{ex:logistic}
Consider binary classification with $y_n \in \{0,1\}$ and logistic link $\mu_n = \sigma(\mathbf{x}_n^\mathsf{T}\boldsymbol\beta)$ where $\sigma(z) = 1/(1+e^{-z})$.
The Fisher weight is $w_n = \mu_n(1-\mu_n)$, maximized at $\mu_n = 0.5$ where $w_n = 0.25$.
For a component with $N^{(\boldsymbol\alpha)}$ observations, the average weight is $\bar{w}^{(\boldsymbol\alpha)} = (N^{(\boldsymbol\alpha)})^{-1} \sum_{n \in \Omega^{(\boldsymbol\alpha)}} \mu_n(1-\mu_n)$.
Substituting into Eq.~\ref{eq:glm-scaling}:
\begin{equation}
    \tau^{(\boldsymbol\alpha)} \leq \frac{1}{\sqrt{2p \cdot \bar{w}^{(\boldsymbol\alpha)} \cdot N^{(\boldsymbol\alpha)}}}
    = \frac{1}{\sqrt{2p \cdot \sum_{n \in \Omega^{(\boldsymbol\alpha)}} \mu_n(1-\mu_n)}}.
\end{equation}
For balanced classes ($\bar{w} \approx 0.25$), the bound gives $\tau \leq 1/\sqrt{2p \cdot 0.25 \cdot N^{(\boldsymbol\alpha)}} = \sqrt{2}/\sqrt{p \cdot N^{(\boldsymbol\alpha)}}$ -- twice the Gaussian case.
For imbalanced data with prevalence $\pi \ll 1$, $\bar{w} \approx \pi(1-\pi) \approx \pi$, yielding $\tau \leq 1/\sqrt{2p \cdot \pi \cdot N^{(\boldsymbol\alpha)}}$.
In a two-level hierarchy (global + group), the group-level prior scales as $\tau_g \leq \sqrt{2}/\sqrt{p \cdot N_g}$ for balanced groups.
\end{example}

\begin{remark}[Application to tree-based models]
\label{rem:trees}
The scaling law applies beyond lattice models to any piecewise regression with $p$ parameters per cell.
For model trees~\citep{QuinlanJR1992} and local linear forests that fit $p$-dimensional linear models per leaf, the constraint $\gamma_{\text{leaf}} = p/N_{\text{leaf}} < 1$ requires minimum leaf sizes scaling with $p$.
The regularization scaling $\tau_{\text{leaf}} \leq \sigma/\sqrt{2p \cdot N_{\text{leaf}}}$ suggests that ridge penalties should adapt to local sample size to maintain a bounded complexity contribution per leaf.
Standard model trees use uniform regularization across leaves; adaptive scaling may improve generalization.
More broadly, hierarchical shrinkage --where leaf estimates pool toward parent node estimates --provides a tree analog of this decomposition, and is implemented in Bayesian approaches like BART~\citep{chipmanBARTBayesianAdditive2010}; for a review of Bayesian shrinkage priors in penalized regression, see \citet{vanerpShrinkagePriorsBayesian2019}.
\end{remark}

\subsection{MAP estimation for scalability.}
While this framework, being a form of multilevel statistical modeling, is most-naturally Bayesian, computing the full posterior (via MCMC or variational inference) incurs significant computational overhead that is impractical for large-scale benchmarking.
This paper therefore uses maximum a posteriori (MAP) estimation: we optimize the penalized negative log-likelihood $-\log f(y|x,\boldsymbol\theta) + \lambda \|\boldsymbol\theta\|^2_2$.
The generalization-preserving bound on $\tau$ (Eq.~\ref{eq:scaling-law}) translates directly to the regularization strength $\lambda \propto \tau^{-2}$, and the effective degrees of freedom retain the same shrinkage interpretation.
The WAIC variance term (Eq.~\ref{eq:waic-prelim}) is approximated via the Laplace approximation around the MAP estimate (Appendix~\ref{app:glm}).
This MAP formulation preserves the theoretical insights while enabling efficient gradient-based optimization.

\subsection{Local stacking}
\label{sec:ensembling}

The mesoscopic structure enables region-specific stacking~\citep{yaoBayesianHierarchicalStacking2021}: different base models can be weighted differently in different regions of feature space, with the weight function itself expressed using a parameter decomposed over a lattice.
Combining base model logits (rather than probabilities) keeps the ensemble within the same model class.

\paragraph{Local model weights by decomposition.}
Given $M$ base models with logit outputs $\eta_1(\mathbf{x}), \ldots, \eta_M(\mathbf{x})$, we parameterize local stacking weights via a decomposed softmax to ensure weights remain non-negative and sum to one:
\begin{equation}
    w_m(\boldsymbol\kappa) = \frac{\exp(v_m^{(\boldsymbol\kappa)})}{\sum_{m'} \exp(v_{m'}^{(\boldsymbol\kappa)})}, \quad
    v_m^{(\boldsymbol\kappa)} = v_m^{()} + \sum_{i} v_m^{(\alpha_i=\kappa_i)} + \sum_{i<j} v_m^{(\alpha_i, \alpha_j)} + \cdots
\end{equation}
where each $v_m$ is a decomposed parameter with the same structure as the base models.
The ensemble prediction is $\eta_{\text{ens}}(\mathbf{x}) = \sum_m w_m(\boldsymbol\kappa(\mathbf{x})) \cdot \eta_m(\mathbf{x})$.
Truncating at order $K{=}1$ yields a weight function with $M(1 + d)$ parameters for a $d$-dimensional lattice.
Learning stacking weights on the same data used to fit base models induces overfitting.
We optimize stacking weights using a leverage-based approximation to leave-one-out (LOO) cross-validation~\citep{cookResidualsInfluenceRegression1982}: for observation $i$ in cell $\boldsymbol\kappa_i$, the leverage is $h_{ii} \approx M / n_{\boldsymbol\kappa_i}$, and the LOO loss is $\mathcal{L}_{\text{LOO}} \approx -\frac{1}{N} \sum_i \frac{1}{1 - h_{ii}} \cdot \ell(y_i, \eta_{\text{ens}}(\mathbf{x}_i))$, where $\ell$ is the log-likelihood evaluated at the ensemble linear predictor.
We minimize $\mathcal{L}_{\text{LOO}}$ with respect to the $v_m$ parameters via gradient descent.

\section{Experiments}

We validate the theoretical predictions through Monte Carlo simulations on synthetic data (Appendix~\ref{app:simulations}) and compare against standard methods on public data from the UCI Machine Learning Repository~\citep{Dua2019}.
All code is available at \url{https://anonymous.4open.science/r/bq-80E9}.

\subsection{UCI benchmark comparisons}

We compare against standard tabular learning methods on eleven classification benchmarks spanning sample sizes from $N=270$ (Heart Disease) to $N=98050$ (HIGGS) and feature dimensions from $p=5$ (Phoneme) to $p=1776$ (Bioresponse).
Table~\ref{tab:uci-comparison} reports test AUC using 5-fold stratified cross-validation.
Our hierarchical model uses generalization-preserving regularization with truncation order determined by the theory-derived SNR criterion (Section~\ref{sec:rg-flow}).
The general procedure for reproducing these models and the exact settings used are presented in Supplemental Section~\ref{app:lattice-construction} and Supplement Table~\ref{tab:hyperparameters}.

\begin{table}[h]
    \centering
    \caption{\textbf{Test AUC on classification benchmarks (mean $\pm$ std, 5-fold stratified CV).}
    Methods: LR = Logistic Regression, MLP = Multi-Layer Perceptron, RF = Random Forest, XGB = XGBoost, LGBM = LightGBM, GAMINet~\citep{yangGamiNetExplainableNeural2021}, EBM = Explainable Boosting Machine~\citep{noriInterpretMLUnifiedFramework2019}.
    Bold = best, underline = second-best. *Did not converge. $^\dagger$Ensemble of 4 models with LOO-weighted stacking; +EBM achieves 0.797.}
    \label{tab:uci-comparison}
    \small
    \begin{tabular}{lcccccccc}
        \hline
        Dataset & LR & MLP & RF & XGB & LGBM & GAMINet & EBM & Ours \\
        \hline
        Heart & .787 & .677 & .744 & .722 & .739 & .604 & \underline{.892} & \textbf{.907} \\
        {\tiny $N{=}270$} & {\tiny $\pm$.056} & {\tiny $\pm$.156} & {\tiny $\pm$.055} & {\tiny $\pm$.055} & {\tiny $\pm$.053} & {\tiny $\pm$.127} & {\tiny $\pm$.031} & {\tiny $\pm$.039} \\[1pt]
        German & .786 & .786 & \textbf{.794} & .771 & .775 & .784 & .773 & \underline{.788} \\
        {\tiny $N{=}1000$} & {\tiny $\pm$.019} & {\tiny $\pm$.027} & {\tiny $\pm$.014} & {\tiny $\pm$.032} & {\tiny $\pm$.020} & {\tiny $\pm$.010} & {\tiny $\pm$.027} & {\tiny $\pm$.021} \\[1pt]
        Madelon & .567 & .596 & .778 & .884 & \underline{.890} & .826 & .832 & \textbf{.905} \\
        {\tiny $N{=}2600$} & {\tiny $\pm$.016} & {\tiny $\pm$.018} & {\tiny $\pm$.011} & {\tiny $\pm$.008} & {\tiny $\pm$.007} & {\tiny $\pm$.014} & {\tiny $\pm$.014} & {\tiny $\pm$.014} \\[1pt]
        Bioresponse & .797 & .848 & \underline{.868} & \textbf{.868} & \underline{.868} & * & .859 & .843 \\
        {\tiny $N{=}3751$} & {\tiny $\pm$.007} & {\tiny $\pm$.009} & {\tiny $\pm$.009} & {\tiny $\pm$.003} & {\tiny $\pm$.007} &  & {\tiny $\pm$.008} & {\tiny $\pm$.010} \\[1pt]
        Spambase & .971 & .981 & .983 & \underline{.988} & .988 & .972 & .987 & \textbf{.994} \\
        {\tiny $N{=}4601$} & {\tiny $\pm$.003} & {\tiny $\pm$.003} & {\tiny $\pm$.005} & {\tiny $\pm$.003} & {\tiny $\pm$.003} & {\tiny $\pm$.003} & {\tiny $\pm$.003} & {\tiny $\pm$.003} \\[1pt]
        Phoneme & .812 & .935 & \textbf{.961} & .954 & \underline{.953} & .922 & .943 & .937 \\
        {\tiny $N{=}5404$} & {\tiny $\pm$.008} & {\tiny $\pm$.005} & {\tiny $\pm$.002} & {\tiny $\pm$.004} & {\tiny $\pm$.002} & {\tiny $\pm$.003} & {\tiny $\pm$.004} & {\tiny $\pm$.005} \\[1pt]
        Taiwan & .723 & .770 & \underline{.780} & .776 & .778 & .780 & \textbf{.784} & .772 \\
        {\tiny $N{=}30000$} & {\tiny $\pm$.004} & {\tiny $\pm$.005} & {\tiny $\pm$.005} & {\tiny $\pm$.005} & {\tiny $\pm$.005} & {\tiny $\pm$.006} & {\tiny $\pm$.007} & {\tiny $\pm$.006} \\[1pt]
        Bank & .890 & .924 & .903 & .906 & .906 & \underline{.932} & \textbf{.934} & .918 \\
        {\tiny $N{=}45211$} & {\tiny $\pm$.003} & {\tiny $\pm$.005} & {\tiny $\pm$.003} & {\tiny $\pm$.003} & {\tiny $\pm$.003} & {\tiny $\pm$.005} & {\tiny $\pm$.004} & {\tiny $\pm$.003} \\[1pt]
        Electricity & .819 & .925 & .913 & .927 & .928 & .896 & \textbf{.959} & \underline{.934} \\
        {\tiny $N{=}45312$} & {\tiny $\pm$.005} & {\tiny $\pm$.005} & {\tiny $\pm$.003} & {\tiny $\pm$.003} & {\tiny $\pm$.002} & {\tiny $\pm$.003} & {\tiny $\pm$.002} & {\tiny $\pm$.002} \\[1pt]
        Adult & .907 & .909 & .911 & \underline{.929} & .929 & .914 & \textbf{.930} & .915 \\
        {\tiny $N{=}48842$} & {\tiny $\pm$.002} & {\tiny $\pm$.002} & {\tiny $\pm$.002} & {\tiny $\pm$.002} & {\tiny $\pm$.002} & {\tiny $\pm$.002} & {\tiny $\pm$.002} & {\tiny $\pm$.002} \\[1pt]
        HIGGS & .682 & .794 & .795 & \underline{.800} & \textbf{.804} & .792 & {.793} & .793$^\dagger$ \\
        {\tiny $N{=}98050$} & {\tiny $\pm$.002} & {\tiny $\pm$.002} & {\tiny $\pm$.003} & {\tiny $\pm$.002} & {\tiny $\pm$.002} & {\tiny $\pm$.002} & {\tiny $\pm$.001} & {\tiny $\pm$.002} \\
        \hline
    \end{tabular}
\end{table}

Our method achieves best or second-best performance on 5 of 11 datasets.
On Madelon ($p{=}500$), the lattice-based feature selection identifies the informative subspace despite many irrelevant features.
On Electricity, a lattice with date, day, hour, and price dimensions captures temporal-price interactions that tree ensembles miss.
On high-dimensional Bioresponse ($p{=}1776$, $N{=}3751$), the features are extremely sparse molecular descriptors with no obvious low-dimensional grouping structure.
Ensembling models with different feature rankings (LR-selected vs RF-selected) via logit averaging achieves 0.843 AUC, narrowing but not closing the gap to tree ensembles (0.868).

On HIGGS, a single model achieves 0.788; ensembling 4 models with LOO-weighted local stacking (Section~\ref{sec:ensembling}) improves this to 0.793.
The ensemble combines diverse architectures: a 4-dimensional order-2 intercept lattice ($\approx5\times10^3$ cells), a 3-dimensional order-3 lattice, an 8-dimensional order-1 lattice ($\approx$1.7M cells), and a high-resolution 1-dimensional lattice (32 bins).
The weight function uses a 3-dimensional order-1 decomposition.

Because EBM is also pairwise additive, it can be included in the ensemble while preserving interpretability.
Ensembling 2 of our models with EBM using the same LOO-weighted stacking achieves 0.797 AUC -- outperforming EBM alone (0.793) by exploiting complementary strengths.
The resulting model remains interpretable: main effects and pairwise interactions from all base models, combined with a learned weight function.

\section{Discussion}

Our method performs particularly well on small to moderate datasets, where the explicit lattice structure and generalization-preserving regularization provide strong inductive bias.
On Heart Disease ($N{=}270$), German Credit ($N{=}1000$), Madelon ($N{=}2600$), and Spambase ($N{=}4601$), our approach outperforms logistic regression and often matches or exceeds tree ensembles.
On larger datasets such as Adult, Bank, and HIGGS, our method remains competitive with tree ensembles but typically trails EBM, which benefits from its boosting approach and automatic interaction detection at scale.
This pattern suggests that the fixed lattice structure is most advantageous when sample sizes are modest relative to the complexity of the decision boundary -- precisely the regime where the generalization-preserving regularization has its strongest effect.

\subsection{Limitations}

\textbf{Methodological.} Unlike adaptive methods (decision trees, neural networks), our approach requires fixing the lattice structure -- which features define the partition and how to bin them.
This manual specification is where domain knowledge enters: scientific questions should be encoded directly into the lattice rather than left implicit in regression coefficients.
For example, if one hypothesizes sex differences in effects, sex becomes a lattice dimension (not merely a covariate); if clinical guidelines define risk categories via BMI or lab value thresholds, those thresholds define natural bin boundaries.
Low-cardinality categoricals serve directly as lattice dimensions, while high-cardinality features (zip codes, diagnosis codes) require grouping -- by domain knowledge, outcome prevalence, or learned embeddings.
Continuous features are discretized via percentile binning subject to the constraint $L < (N/p)^{1/d}$ (Remark~\ref{rem:bin-count}).
Large language models can assist in exploring this design space (Appendix~\ref{app:lattice-construction}).
While some may view this manual labor as a weakness, we maintain that encoding structure explicitly is preferable to arguing posthoc that it exists in a black-box model.

The number of cells grows as $L^d$ where $L$ is bins per dimension and $d$ is the number of lattice dimensions.
Dense high-dimensional lattices are impractical even with modest $L$.
The bin count constraint mitigates this by forcing smaller $L$ as $d$ increases, but fundamentally limits the resolution achievable with many factors.
Sparsity-inducing priors (e.g., Horseshoe, spike-and-slab) could potentially address this by automatically pruning empty or irrelevant cells, though extending the replica analysis to such priors remains future work.

\textbf{Analytical.}
The theoretical results in this paper -- the scaling law (Eq.~\ref{eq:scaling-law}), bin count constraint (Eq.~\ref{eq:bin-constraint}), and critical order (Eq.~\ref{eq:critical-order}) -- are based on approximations (Laplace approximation, replica-symmetric saddle point) and describe expected behavior averaged over data realizations.
They are not exact bounds: a particular dataset may benefit from regularization stronger or weaker than Eq.~\ref{eq:scaling-law} suggests.
Practitioners should treat these as principled starting points for hyperparameter search, not as final values.
Cross-validation remains the gold standard for tuning; the theory narrows the search space and provides interpretable defaults.

While we extend the analysis to generalized linear models (Section~\ref{sec:glm-extension}), the effective degrees of freedom depends on the Fisher information evaluated at the posterior mean.
For severely misspecified models or extreme class imbalance, the local quadratic approximation underlying the scaling law may be inaccurate.

The replica calculation assumes replica symmetry -- that all replicas are statistically equivalent.
This assumption hold asymptotically when the posterior has a single dominant mode.
For highly multimodal posteriors, replica symmetry breaking may occur and our mean-field predictions become inaccurate, though the regularization scheme remains applicable as a heuristic.

The critical order $K^*$ (Eq.~\ref{eq:critical-order}) depends on the effect decay rate $\rho$, which is unknown in practice.
One approach is to estimate $\rho$ empirically: fit models at orders $K=0,1,2$ and compute the ratio of estimated effect variances $\hat\rho = \text{Var}[\hat\theta^{(1)}]/\text{Var}[\hat\theta^{(0)}]$.
Alternatively, the generalization-preserving regularization sidesteps this issue by ensuring safe expansion regardless of the true $\rho$: if effects decay faster than assumed, the prior shrinks spurious terms to zero; if slower, the prior permits their estimation.

\subsection{Future directions}

\textbf{Training optimizations.}
One promising direction is to learn the binned dimensions that define the lattice structure from data by clustering in a latent representation space -- for example, quantizing an autoencoder's latent codes to define hierarchical groupings. Techniques used in EBMs and other architectures such as heuristics for bin modifications would be useful to implement.
Other techniques such as boosting and bagging also may be useful efficiently finding accurate models.

\textbf{Continual learning.}
Models of this class are suitable for continual learning, where the lattice is progressively refined as data accumulates -- starting coarse, then expanding to finer partitions only where the data supports additional complexity, guided by the generalization-preserving bounds.

\section{Conclusion}

We have presented a framework for piecewise GLMs with hierarchical parameter decomposition, drawing on ideas from renormalization group theory and replica analysis.
The central contribution is a generalization-preserving regularization scheme (Eq.~\ref{eq:scaling-law}) that allows model complexity to grow without the usual bias-variance penalty: components with insufficient data support are automatically shrunk, while those with signal are retained.
The replica calculation provides complementary guidance on lattice resolution (Remark~\ref{rem:bin-count}), ensuring the mean-field approximation remains valid.

These results bridge interpretable modeling through classical multilevel regression modeling and modern scalable architectures.
The explicit lattice structure enables direct inspection of which feature combinations drive predictions, while the scaling laws provide guardrails against overfitting.
We hope this work 1) helps rejuvenate the desire for using intrinsically interpretable models over blackbox methods and 2) encourages further development at the intersection of statistical physics and machine learning, particularly for applications where understanding model behavior is as important as predictive accuracy.

\begin{ack}

This research was supported [in part] by the Intramural Research Program of the National Institutes of Health (NIH). The contributions of the NIH author(s) are considered Works of the United States Government. The findings and conclusions presented in this paper are those of the author(s) and do not necessarily reflect the views of the NIH or the U.S. Department of Health and Human Services.

The author thanks Carson Chow, Hongjing Xia, Shashaank Vattikuti, Patrick Fletcher, and the members of the ``mederrata'' CMS AI Health innovations challenge team.
\end{ack}
%
\newpage

\bibliographystyle{plainnat}
\bibliography{mederrata}

\newpage
\appendix

\section{Supplementary Material}

\subsection{Derivation of the generalization-preserving scaling law}
\label{app:generalization-preserving}

We derive the scaling law (Eq.~\ref{eq:scaling-law}) from first principles.
Consider first a single parameter $\theta^{(\boldsymbol\alpha)}$ in a Gaussian model with noise variance $\sigma^2$.
The parameter is observed through $N^{(\boldsymbol\alpha)}$ observations with mean $\bar{y}^{(\boldsymbol\alpha)}$.

The prior is $\theta^{(\boldsymbol\alpha)} \sim \mathcal{N}(0, \tau^2)$ where we write $\tau = \tau^{(\boldsymbol\alpha)}$ for brevity.
The posterior is
\begin{align}
    \theta^{(\boldsymbol\alpha)} | \bar{y}^{(\boldsymbol\alpha)}
    &\sim \mathcal{N}\left(
        \frac{N^{(\boldsymbol\alpha)} \tau^2}{N^{(\boldsymbol\alpha)} \tau^2 + \sigma^2} \bar{y}^{(\boldsymbol\alpha)},
        \frac{\tau^2 \sigma^2 / N^{(\boldsymbol\alpha)}}{\tau^2 + \sigma^2 / N^{(\boldsymbol\alpha)}}
    \right) \nonumber \\
    &= \mathcal{N}\left( s \cdot \bar{y}^{(\boldsymbol\alpha)}, \frac{s \sigma^2}{N^{(\boldsymbol\alpha)}} \right),
    \label{eq:app-posterior}
\end{align}
where the shrinkage factor $s$ appearing in Eq.~\ref{eq:app-posterior} is
\begin{equation}
    s = \frac{N^{(\boldsymbol\alpha)} \tau^2}{N^{(\boldsymbol\alpha)} \tau^2 + \sigma^2}.
    \label{eq:app-shrinkage}
\end{equation}

The WAIC complexity penalty for a parameter equals its effective degrees of freedom, which for ridge regression is
\begin{equation}
    \text{df}_{\text{eff}} = \text{tr}\left[ (\mathbf{X}^\mathsf{T}\mathbf{X} + \sigma^2\tau^{-2}\mathbf{I})^{-1} \mathbf{X}^\mathsf{T}\mathbf{X} \right].
    \label{eq:app-df}
\end{equation}
Specializing Eq.~\ref{eq:app-df} to a single parameter observed through $N^{(\boldsymbol\alpha)}$ observations (i.e., $\mathbf{X} = \mathbf{1}$, so $\mathbf{X}^\mathsf{T}\mathbf{X} = N^{(\boldsymbol\alpha)}$):
\begin{equation}
    \text{df}_{\text{eff}} = \frac{N^{(\boldsymbol\alpha)}}{N^{(\boldsymbol\alpha)} + \sigma^2/\tau^2} = \frac{N^{(\boldsymbol\alpha)} \tau^2}{N^{(\boldsymbol\alpha)} \tau^2 + \sigma^2} = s.
    \label{eq:app-df-single}
\end{equation}
Thus Eq.~\ref{eq:app-df-single} shows that effective degrees of freedom equals the shrinkage factor from Eq.~\ref{eq:app-shrinkage}: a fully shrunk parameter ($s \to 0$) contributes no complexity, while an unregularized parameter ($s \to 1$) contributes one full degree of freedom.

The generalization-preserving condition requires that adding this parameter not increase expected WAIC.
Since WAIC $\approx$ training loss $+ 2 \cdot \text{df}_{\text{eff}}$, we need the reduction in training loss to exceed $2 \cdot \text{df}_{\text{eff}}$.
When the true effect is zero, training loss reduction averages $\text{df}_{\text{eff}} \cdot \sigma^2 / N^{(\boldsymbol\alpha)}$, which falls short of $2 \cdot \text{df}_{\text{eff}}$ unless $\text{df}_{\text{eff}}$ is small.
Requiring $\text{df}_{\text{eff}} \leq 1/2$ ensures that even null parameters do not hurt generalization:
\[
    \frac{N^{(\boldsymbol\alpha)} \tau^2}{N^{(\boldsymbol\alpha)} \tau^2 + \sigma^2} \leq \frac{1}{2}
    \quad \Rightarrow \quad
    \tau \leq \frac{\sigma}{\sqrt{N^{(\boldsymbol\alpha)}}}.
\]

For a component with $p$ regression coefficients, two bounding strategies are possible:
\begin{enumerate}
    \item \textbf{Per-parameter:} Use $\tau = \sigma/\sqrt{N^{(\boldsymbol\alpha)}}$ for each coefficient, bounding total $\text{df}_{\text{eff}} \leq p/2$.
    \item \textbf{Per-component:} Use $\tau = \sigma/\sqrt{2p \cdot N^{(\boldsymbol\alpha)}}$ for all coefficients, bounding total $\text{df}_{\text{eff}} \leq 1/2$.
\end{enumerate}
The main text adopts per-component bounding (Eq.~\ref{eq:scaling-law}), which provides more aggressive regularization for high-dimensional interaction terms.

\subsection{Extension to generalized linear models}
\label{app:glm}

The Gaussian derivation above assumes constant noise variance $\sigma^2$.
For classification and count data, the noise structure depends on the mean, requiring a modified analysis.
This section derives the generalization-preserving bound for GLMs (Eq.~\ref{eq:glm-scaling} in the main text).

For a GLM with canonical link, the negative log-likelihood is
\begin{equation}
    -\ell(\boldsymbol\beta) = \sum_{n=1}^N \left[ b(\eta_n) - y_n \eta_n \right] / a(\phi) - c(y_n, \phi),
    \label{eq:app-glm-nll}
\end{equation}
where $\eta_n = \mathbf{x}_n^\mathsf{T}\boldsymbol\beta$.
The Hessian of Eq.~\ref{eq:app-glm-nll} is $\mathbf{H} = \mathbf{X}^\mathsf{T} \mathbf{W} \mathbf{X} / a(\phi)$ where $W_{nn} = b''(\eta_n)$.

For logistic regression, $b(\eta) = \log(1 + e^\eta)$, so $b'(\eta) = \sigma(\eta)$ (the sigmoid) and $b''(\eta) = \sigma(\eta)(1 - \sigma(\eta))$.
The dispersion is $a(\phi) = 1$.

The posterior under a Gaussian prior $\boldsymbol\beta \sim \mathcal{N}(0, \tau^2 \mathbf{I})$ is approximated by a Laplace approximation:
\begin{equation}
    \boldsymbol\beta | \mathbf{y} \approx \mathcal{N}\left( \hat{\boldsymbol\beta}, \left( \mathbf{X}^\mathsf{T} \hat{\mathbf{W}} \mathbf{X} + \tau^{-2}\mathbf{I} \right)^{-1} \right),
    \label{eq:app-glm-posterior}
\end{equation}
where $\hat{\mathbf{W}}$ is evaluated at the MAP estimate $\hat{\boldsymbol\beta}$.

From Eq.~\ref{eq:app-glm-posterior}, the effective degrees of freedom is
\begin{equation}
    \text{df}_{\text{eff}} = \text{tr}\left[ \left( \mathbf{X}^\mathsf{T} \hat{\mathbf{W}} \mathbf{X} + \tau^{-2}\mathbf{I} \right)^{-1} \mathbf{X}^\mathsf{T} \hat{\mathbf{W}} \mathbf{X} \right].
    \label{eq:app-glm-df}
\end{equation}

Specializing Eq.~\ref{eq:app-glm-df} to a single parameter with $N$ observations having average weight $\bar{w} = N^{-1}\sum_n w_n$:
\begin{equation}
    \text{df}_{\text{eff}} = \frac{N \bar{w} \tau^2}{N \bar{w} \tau^2 + 1}.
    \label{eq:app-glm-df-single}
\end{equation}
Bounding Eq.~\ref{eq:app-glm-df-single} by $\text{df}_{\text{eff}} \leq 1/2$ and solving for $\tau$ yields
\begin{equation}
    \tau \leq \frac{1}{\sqrt{N \bar{w}}}.
    \label{eq:app-glm-bound}
\end{equation}

For binary classification with class probability $\pi$, the average weight is $\bar{w} = \pi(1-\pi)$.
At $\pi = 0.5$ (balanced), $\bar{w} = 0.25$, so Eq.~\ref{eq:app-glm-bound} gives $\tau \leq 2/\sqrt{N}$.
At $\pi = 0.1$ (imbalanced), $\bar{w} = 0.09$, so $\tau \leq 3.3/\sqrt{N}$ -- the regularization is weaker to account for the lower effective sample size.

As in the Gaussian case, for a component with $p$ regression coefficients, the per-component bound uses $\tau = 1/\sqrt{2p \cdot N \bar{w}}$ to ensure total $\text{df}_{\text{eff}} \leq 1/2$.
In practice, we iteratively update the weights and regularization during optimization.
Starting from $\bar{w}^{(0)} = 0.25$ (balanced initialization), at iteration $t$ we compute
\begin{equation}
    \bar{w}^{(t)} = \frac{1}{N} \sum_{n=1}^N \hat{\mu}_n^{(t)}(1 - \hat{\mu}_n^{(t)}), \quad
    \tau^{(t)} = \frac{1}{\sqrt{2p \cdot N \cdot \bar{w}^{(t)}}},
    \label{eq:app-glm-iterative}
\end{equation}
where $\hat{\mu}_n^{(t)} = \sigma(\mathbf{x}_n^\mathsf{T}\hat{\boldsymbol\beta}^{(t)})$.
Eq.~\ref{eq:app-glm-iterative} adapts the per-component bound to the current class balance and converges in practice within a few outer iterations.

\subsection{WAIC variance calculation for Gaussian regression}
\label{app:waic-variance}

The WAIC (Eq.~\ref{eq:waic-prelim}) includes a variance term that penalizes model complexity.
This section derives that term explicitly for the conjugate Normal-Inverse-Gamma model, showing how it depends on leverage and residuals.
This derivation is not required for the main results but provides intuition for how WAIC captures overfitting risk.

We derive the variance term for the conjugate Normal-Inverse-Gamma model.
The log-likelihood for observation $n$ is
\begin{equation}
    \ell_n = -\frac{1}{2}\log(2\pi\sigma^2) - \frac{(y_n - \mathbf{x}_n\boldsymbol\beta)^2}{2\sigma^2}.
    \label{eq:app-loglik}
\end{equation}

Applying the law of total variance to Eq.~\ref{eq:app-loglik} with respect to the posterior $\pi(\boldsymbol\beta, \sigma^2 | \mathbf{y}, \mathbf{X})$:
\begin{align}
    \text{Var}[\ell_n]
    &= \mathbb{E}_{\sigma^2}\left[ \text{Var}_{\boldsymbol\beta|\sigma^2}[\ell_n] \right]
     + \text{Var}_{\sigma^2}\left[ \mathbb{E}_{\boldsymbol\beta|\sigma^2}[\ell_n] \right].
    \label{eq:app-total-var}
\end{align}
We evaluate each term in Eq.~\ref{eq:app-total-var} separately.

For the inner variance (conditional on $\sigma^2$), applying Eq.~\ref{eq:app-loglik}:
\begin{align}
    \text{Var}_{\boldsymbol\beta|\sigma^2}[\ell_n]
    &= \text{Var}_{\boldsymbol\beta|\sigma^2}\left[ -\frac{(y_n - \mathbf{x}_n\boldsymbol\beta)^2}{2\sigma^2} \right] \nonumber \\
    &= \frac{1}{4\sigma^4} \text{Var}_{\boldsymbol\beta|\sigma^2}\left[ (y_n - \mathbf{x}_n\boldsymbol\beta)^2 \right].
    \label{eq:app-inner-var}
\end{align}

Let $\hat{r}_n = y_n - \mathbf{x}_n\hat{\boldsymbol\beta}$ be the posterior mean residual and $h_n = \mathbf{x}_n\hat\Sigma^{-1}\mathbf{x}_n^\mathsf{T}$ the leverage.
Expanding $(y_n - \mathbf{x}_n\boldsymbol\beta)^2 = (\hat{r}_n - \mathbf{x}_n(\boldsymbol\beta - \hat{\boldsymbol\beta}))^2$ in Eq.~\ref{eq:app-inner-var}:
\begin{equation}
    \text{Var}_{\boldsymbol\beta|\sigma^2}\left[ (y_n - \mathbf{x}_n\boldsymbol\beta)^2 \right]
    = 4\hat{r}_n^2 \cdot \sigma^2 h_n + 2\sigma^4 h_n^2,
    \label{eq:app-quadratic-var}
\end{equation}
using the variance of a quadratic form in a Gaussian.
Substituting Eq.~\ref{eq:app-quadratic-var} into Eq.~\ref{eq:app-inner-var}:
\begin{equation}
    \text{Var}_{\boldsymbol\beta|\sigma^2}[\ell_n]
    = \frac{\hat{r}_n^2 h_n}{\sigma^2} + \frac{h_n^2}{2}.
    \label{eq:app-inner-result}
\end{equation}

For the outer variance in Eq.~\ref{eq:app-total-var}, the inner expectation is
\begin{equation}
    \mathbb{E}_{\boldsymbol\beta|\sigma^2}[\ell_n]
    = -\frac{1}{2}\log(2\pi\sigma^2) - \frac{\hat{r}_n^2 + \sigma^2 h_n}{2\sigma^2}.
    \label{eq:app-inner-exp}
\end{equation}
The variance of Eq.~\ref{eq:app-inner-exp} over $\sigma^2 \sim \text{InvGamma}(a_N, b_N)$ involves the moments
\begin{align}
    \text{Var}_{\sigma^2}[\log\sigma^2] &= \psi'(a_N), \nonumber \\
    \text{Var}_{\sigma^2}[1/\sigma^2] &= \frac{a_N}{b_N^2}, \nonumber \\
    \text{Cov}_{\sigma^2}[\log\sigma^2, 1/\sigma^2] &= -\frac{1}{b_N}.
    \label{eq:app-ig-moments}
\end{align}

Applying Eq.~\ref{eq:app-ig-moments} to compute the variance of Eq.~\ref{eq:app-inner-exp}:
\begin{align}
    \text{Var}_{\sigma^2}\left[ \mathbb{E}_{\boldsymbol\beta|\sigma^2}[\ell_n] \right]
    &= \frac{\psi'(a_N)}{4}
     + \frac{a_N \hat{r}_n^4}{4b_N^2}
     - \frac{\hat{r}_n^2}{2b_N}.
    \label{eq:app-outer-var}
\end{align}

Finally, combining Eq.~\ref{eq:app-inner-result} and Eq.~\ref{eq:app-outer-var} via Eq.~\ref{eq:app-total-var}:
\begin{equation}
    \text{Var}[\ell_n]
    = \frac{a_N \hat{r}_n^2 h_n}{b_N} + \frac{h_n^2}{2}
    + \frac{\psi'(a_N)}{4} + \frac{a_N \hat{r}_n^4}{4b_N^2}
    - \frac{\hat{r}_n^2}{2b_N},
    \label{eq:app-total-result}
\end{equation}
Eq.~\ref{eq:app-total-result} gives the variance term in the WAIC (Eq.~\ref{eq:waic-prelim}) for the conjugate case.

\subsection{Replica calculation details}
\label{app:replica}

The main text claims that effective degrees of freedom (Eq.~\ref{eq:app-df-single}) controls generalization.
This section verifies that claim using the replica method from statistical physics, which computes expected generalization error by averaging over random design matrices.
The calculation confirms that $\text{df}_{\text{eff}}$ emerges as the natural complexity measure and recovers the classical result of~\citet{kroghGeneralizationLinearPerceptron1992}.

The replica method computes $\mathbb{E}[\log Z]$ via the identity $\log Z = \lim_{n \to 0}(Z^n - 1)/n$.
We introduce $n$ ``replicas'' -- independent copies of the parameter vector $\boldsymbol\beta^a$ for $a = 1, \ldots, n$ -- and compute $\mathbb{E}[Z^n]$ before taking $n \to 0$.

\paragraph{Averaging over the design matrix.}
The replicated partition function before averaging is
\begin{equation}
    Z^n = \int \exp\left( -\sum_{a=1}^n \frac{\|\mathbf{y} - \mathbf{X}\boldsymbol\beta^a\|^2}{2\sigma^2}
                          - \sum_{a=1}^n \frac{\|\boldsymbol\beta^a\|^2}{2\lambda^2} \right)
          \prod_{a=1}^n \dd\boldsymbol\beta^a.
    \label{eq:app-Zn}
\end{equation}

Expanding the quadratic in Eq.~\ref{eq:app-Zn}:
\begin{align}
    \sum_a \|\mathbf{y} - \mathbf{X}\boldsymbol\beta^a\|^2
    &= n\|\mathbf{y}\|^2 - 2\sum_a \mathbf{y}^\mathsf{T}\mathbf{X}\boldsymbol\beta^a
     + \sum_a (\boldsymbol\beta^a)^\mathsf{T}\mathbf{X}^\mathsf{T}\mathbf{X}\boldsymbol\beta^a \nonumber \\
    &= n\|\mathbf{y}\|^2 - 2\sum_a \mathbf{y}^\mathsf{T}\mathbf{X}\boldsymbol\beta^a
     + \sum_{a,b} \delta_{ab} (\boldsymbol\beta^a)^\mathsf{T}\mathbf{X}^\mathsf{T}\mathbf{X}\boldsymbol\beta^b.
    \label{eq:app-expand}
\end{align}

For random design with $X_{ni} \sim \mathcal{N}(0, 1/N)$ i.i.d., averaging Eq.~\ref{eq:app-expand} gives $\mathbb{E}[\mathbf{X}^\mathsf{T}\mathbf{X}] = \mathbf{I}_p$ and the cross-term $\mathbf{y}^\mathsf{T}\mathbf{X}\boldsymbol\beta^a$ has mean zero and variance $\|\mathbf{y}\|^2 \|\boldsymbol\beta^a\|^2 / N$.

The key step is introducing the overlap matrix $Q_{ab} = \frac{1}{p}\boldsymbol\beta^a \cdot \boldsymbol\beta^b$ via the Hubbard-Stratonovich identity:
\begin{equation}
    1 = \int \delta\left(pQ_{ab} - \boldsymbol\beta^a \cdot \boldsymbol\beta^b\right) \dd Q_{ab} \propto \int \exp\left( \frac{1}{2}\hat{Q}_{ab}(\boldsymbol\beta^a \cdot \boldsymbol\beta^b - pQ_{ab}) \right) \dd Q_{ab} \dd\hat{Q}_{ab}.
    \label{eq:app-HS}
\end{equation}
The transformation decouples the replicas: the $\boldsymbol\beta^a$ integrals now factorize, each depending only on $Q_{ab}$ rather than on other replicas directly.
The resulting saddle-point equations yield the free energy (Eq.~\ref{eq:app-free-energy}), from which we extract the effective degrees of freedom (Eq.~\ref{eq:app-df-replica}).

\paragraph{Saddle-point equations.}
Inserting Eq.~\ref{eq:app-HS} into Eq.~\ref{eq:app-Zn} and assuming replica symmetry ($Q_{ab} = q$ for $a \neq b$, $Q_{aa} = Q$), the saddle-point equations are
\begin{align}
    \frac{\partial}{\partial Q}\left[ -\frac{n}{2}Q - \frac{np(n-1)}{2}q
        + n\log\int e^{-\frac{1}{2\lambda^2}\|\boldsymbol\beta\|^2 + \hat{Q}\|\boldsymbol\beta\|^2/2} \dd\boldsymbol\beta \right] &= 0, \nonumber \\
    \frac{\partial}{\partial q}\left[\text{same}\right] &= 0.
    \label{eq:app-saddle}
\end{align}

The Gaussian integral in Eq.~\ref{eq:app-saddle} evaluates to
\begin{equation}
    \int e^{-\frac{1}{2}(\lambda^{-2} - \hat{Q})\|\boldsymbol\beta\|^2} \dd\boldsymbol\beta
    = \left( \frac{2\pi}{\lambda^{-2} - \hat{Q}} \right)^{p/2}.
    \label{eq:app-gaussian}
\end{equation}

Substituting Eq.~\ref{eq:app-gaussian} into Eq.~\ref{eq:app-saddle}, taking derivatives, and solving:
\begin{align}
    Q &= \frac{\lambda^2}{1 - \lambda^2\hat{Q}}, \nonumber \\
    q &= \frac{\lambda^4 \hat{q}}{(1 - \lambda^2\hat{Q})^2},
    \label{eq:app-Qq}
\end{align}
where $\hat{Q}$ and $\hat{q}$ are determined by the conjugate equations.

\paragraph{Free energy.}
Solving Eq.~\ref{eq:app-Qq} and taking $n \to 0$, the quenched free energy density is
\begin{align}
    f &= \lim_{n\to 0} \frac{1}{n}\left( \overline{Z^n} - 1 \right) \nonumber \\
      &= -\frac{1}{2}\log(2\pi\sigma^2) - \frac{1}{2}
       - \frac{\gamma}{2}\log\left(1 + \frac{N\lambda^2}{\sigma^2}\right)
       + \frac{\gamma}{2}\frac{N\lambda^2/\sigma^2}{1 + N\lambda^2/\sigma^2},
    \label{eq:app-free-energy}
\end{align}
where $\gamma = p/N$.

From Eq.~\ref{eq:app-free-energy}, the effective degrees of freedom is the coefficient of the log term:
\begin{equation}
    \text{df}_{\text{eff}} = p \cdot \frac{N\lambda^2}{N\lambda^2 + \sigma^2},
    \label{eq:app-df-replica}
\end{equation}
a classical result~\citep{kroghGeneralizationLinearPerceptron1992}, matching the direct calculation in Eq.~\ref{eq:app-df-single}.

\subsection{Critical phenomena near the fixed point}
\label{app:critical}

The optimal truncation order $K^*$ (Eq.~\ref{eq:critical-order}) marks the transition from underfitting to overfitting.
This section analyzes the sharpness of that transition: how sensitive is generalization to choosing $K$ slightly above or below $K^*$?
The analysis reveals that the transition sharpness depends on the effect decay rate $\rho$ -- when effects decay slowly ($\rho$ close to $L$), the optimal truncation is less sharply defined.

Taylor-expanding the generalization gap near $K^*$:
\begin{equation}
    \Delta S_K \approx \Delta S'_{K^*}(K - K^*) + \frac{1}{2}\Delta S''_{K^*}(K - K^*)^2 + O((K-K^*)^3).
    \label{eq:app-deltaS-taylor}
\end{equation}

For effect sizes $(\theta^{(k)}_*)^2 \sim \rho^k$, the SNR at order $K$ is
\begin{equation}
    \text{SNR}^{(K)} = \frac{N}{L^K} \cdot \rho^K \cdot \frac{1}{\sigma^2}
    = \frac{N\rho^K}{\sigma^2 L^K}.
    \label{eq:app-snr}
\end{equation}

At the fixed point, $\text{SNR}^{(K^*)} = 1$ in Eq.~\ref{eq:app-snr} (the critical case with shrinkage $s=1$). Solving for $K^*$:
\begin{equation}
    K^* = \frac{\log(N/\sigma^2)}{\log(L/\rho)}.
    \label{eq:app-Kstar}
\end{equation}

To evaluate Eq.~\ref{eq:app-deltaS-taylor}, we compute the derivative of the generalization gap:
\begin{align}
    \Delta S'_K &= \frac{\partial}{\partial K}\left[
        -\binom{d}{K}L^K \cdot \text{SNR}^{(K)} \cdot s + \text{df}_{\text{eff}}^{(K)}
    \right] \nonumber \\
    &\approx \binom{d}{K^*}L^{K^*} \cdot \log(L/\rho) \cdot \frac{1}{2},
    \label{eq:app-deltaS-deriv}
\end{align}
where we used Eq.~\ref{eq:app-Kstar} to set $\text{SNR}^{(K^*)} = 1$ and $\text{df}_{\text{eff}}^{(K^*)} = \binom{d}{K^*}L^{K^*}/2$.

From Eq.~\ref{eq:app-deltaS-deriv}, the correlation length (in units of scale) is
\begin{equation}
    \xi = \frac{1}{\Delta S'_{K^*}}
        = \frac{2}{\binom{d}{K^*}L^{K^*}\log(L/\rho)}.
    \label{eq:app-xi}
\end{equation}

For $\rho \ll L$, Eq.~\ref{eq:app-xi} shows the correlation length is short: the transition from underfitting to overfitting occurs over a narrow range of scales.
For $\rho$ approaching $L$, the correlation length diverges: the transition is gradual and the optimal truncation is less sharply defined.

The susceptibility (variance in the optimal truncation across data realizations) scales with the correlation length from Eq.~\ref{eq:app-xi} as
\begin{equation}
    \chi \sim \xi^{2-\eta}
    \label{eq:app-chi}
\end{equation}
for some critical exponent $\eta$.
In the mean-field approximation (replica-symmetric saddle point), $\eta = 0$ and Eq.~\ref{eq:app-chi} gives $\chi \sim \xi^2$.

\subsection{Simulation Studies}
\label{app:simulations}

We validate the theoretical predictions through Monte Carlo simulations.

\begin{figure}[h]
    \centering
    \includegraphics[width=0.75\linewidth]{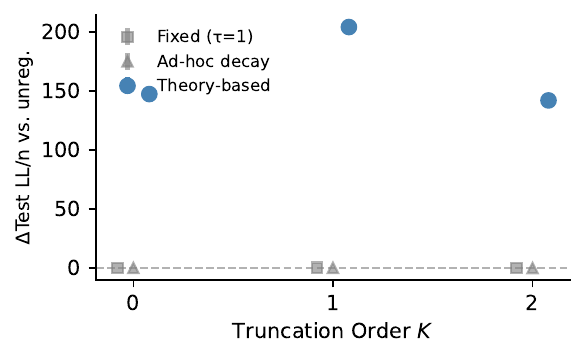}
    \caption{\textbf{Generalization-preserving regularization validation.}
    Improvement in test log-likelihood per observation relative to unregularized baseline across truncation orders $K \in \{0, 1, 2\}$.
    Positive values indicate better generalization.
    The theory-based scheme (blue circles) achieves 147--204 LL units improvement; fixed and ad-hoc decay (gray) show no improvement.}
    \label{fig:gen-preserving}
\end{figure}

\paragraph{Generalization-preserving regularization.}
We generate data from a hierarchical Gaussian model with $d=3$ grouping factors, $L=4$ levels per factor, and $N=10000$ total observations.
True effects decay geometrically across scales: $(\theta^{(k)}_*)^2 \sim \rho^k$ with $\rho = 0.3$.
We fit models truncated at orders $K = 0, 1, 2$ and compare four regularization schemes: (i) unregularized, (ii) fixed ($\tau = 1$), (iii) ad-hoc decay ($\tau^{(k)} = 5 \cdot 0.9^k$), and (iv) generalization-preserving ($\tau^{(k)} = \sigma/\sqrt{2p \cdot N^{(k)}}$).
The generalization-preserving scheme achieves the best test log-likelihood at every truncation order, with improvements of 147--204 log-likelihood units per observation (Figure~\ref{fig:gen-preserving}).

\begin{figure}[h]
    \centering
    \includegraphics[width=0.85\linewidth]{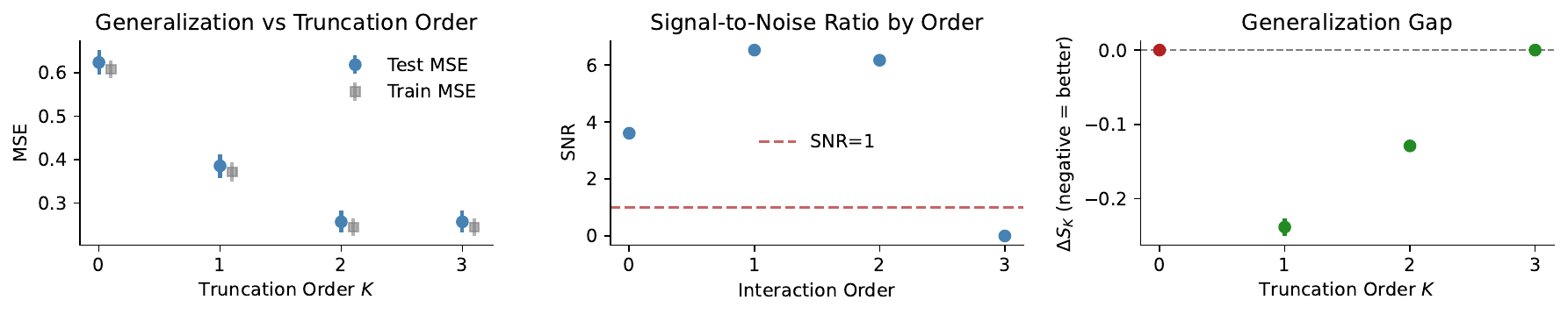}
    \caption{\textbf{RG flow verification using fitted model parameters.}
    Left: Test MSE (lower is better) decreases with truncation order; $K=2$ achieves optimal generalization.
    Center: SNR by interaction order; pairwise effects (order 2) have higher SNR than main effects.
    Right: Generalization gap $\Delta S_K$ (negative is good) confirms each order up to $K=2$ improves generalization.}
    \label{fig:rg-flow}
\end{figure}

\paragraph{RG flow and fixed point.}
Figure~\ref{fig:rg-flow} shows empirical validation using parameters fitted to the German Credit dataset.
The generalization gap $\Delta S_K$ is negative for $K \leq 2$, confirming that each additional interaction order improves out-of-sample performance when underlying effects are present.
Test MSE decreases from 0.62 ($K{=}0$) to 0.39 ($K{=}1$) to 0.26 ($K{=}2$) across 100 replications.

\subsection{Computational Cost Analysis}
\label{app:computational}

Our method computes the coefficient vector by summing over all active decomposition components:
\begin{equation}
    \boldsymbol\beta^{(\boldsymbol\kappa)} = \boldsymbol\beta^{()} + \sum_{i=1}^d \boldsymbol\beta^{(\alpha_i=\kappa_i)} + \sum_{i<j} \boldsymbol\beta^{(\alpha_i=\kappa_i, \alpha_j=\kappa_j)} + \cdots
\end{equation}
Truncation at order $K$ requires summing $C_K = \sum_{k=0}^K \binom{d}{k}$ component tensors, each of dimension $p$.
Each component lookup is $O(1)$ via precomputed indices, yielding total complexity $O(C_K \cdot p)$.
For order-1 truncation, $C_1 = 1 + d$; for order-2, $C_2 = 1 + d + \binom{d}{2}$.

Our method scales linearly in sample size $N$ and feature dimension $p$, with the lattice complexity depending only on $d$ (lattice dimensions) and $K$ (truncation order).

\paragraph{Ensembling with EBM.}
Explainable Boosting Machines are pairwise additive models -- they learn main effects plus selected pairwise interactions via gradient boosting.
This structure is compatible with our model class: the sum of an EBM's logit and our model's logit remains a piecewise additive function with main effects and low-order interactions.
Consequently, one can ensemble our models with EBM using the same local weighting approach described in Section~\ref{sec:ensembling}, combining base model logits (not probabilities) under a learned weight function.
On HIGGS, ensembling four of our models using a 3-dimensional weight lattice with LOO-weighted stacking achieves 0.793 AUC -- outperforming the best single model (0.788).
Including EBM in the ensemble further improves to 0.797 AUC, beating EBM alone (0.793).
The ensemble remains interpretable: it is a weighted sum of pairwise-additive base models, with the weights themselves computed via a low-dimensional decomposition.

For batch inference, our method parallelizes efficiently: the component lookups are embarrassingly parallel across samples via tensor gather operations, and the subsequent summation and dot products leverage optimized BLAS routines.

\paragraph{Timing comparison.}
Table~\ref{tab:timing} reports training and inference times on a 16-thread AMD Ryzen CPU.
Our method trains via iterative gradient descent (500 steps shown); total training time scales linearly with iterations.
We have not tuned learning rates, convergence criteria, or early stopping---models may converge well before 500 steps, so training time could likely be reduced substantially.
Our method is GPU-native (implemented in JAX) and supports minibatch training, enabling efficient scaling to larger datasets.
Inference time is comparable to tree ensembles, reflecting the tensor gather and dot product operations.

\begin{table}[h]
\centering
\small
\caption{\textbf{Training and inference time comparison.} German Credit ($N{=}800$, $p{=}48$) and Adult ($N{=}39073$, $p{=}97$). Times in seconds (training) and milliseconds (inference per batch). Our method uses 500 gradient steps.}
\label{tab:timing}
\begin{tabular}{lcccc}
\toprule
 & \multicolumn{2}{c}{German Credit} & \multicolumn{2}{c}{Adult} \\
\cmidrule(lr){2-3} \cmidrule(lr){4-5}
Method & Train (s) & Inf (ms) & Train (s) & Inf (ms) \\
\midrule
LR & 0.004 & 0.07 & 0.46 & 1.0 \\
MLP & 0.08 & 0.11 & 2.7 & 3.5 \\
RF & 0.11 & 25.0 & 0.48 & 31.2 \\
XGBoost & 0.07 & 0.45 & 0.12 & 1.6 \\
LightGBM & 0.06 & 0.45 & 0.16 & 4.7 \\
EBM & 5.8 & 0.65 & 85.6 & 13.3 \\
Ours & 14$^\dagger$ & 16 & 35$^\dagger$ & 40$^\ast$ \\
\bottomrule
\multicolumn{5}{l}{\footnotesize $^\dagger$500 gradient steps. $^\ast$Estimated from scaling.}
\end{tabular}
\end{table}

\subsection{Lattice Construction Guide}
\label{app:lattice-construction}

Constructing an effective lattice requires balancing statistical constraints with domain knowledge.
We recommend the following workflow.

\paragraph{Step 0: Encode scientific questions directly.}
Before considering predictive features, identify the scientific questions you want the model to answer.
If you want to test whether there are sex differences in outcomes, include sex as a lattice dimension -- not merely as a regression covariate.
If you hypothesize that treatment effects vary by age group, make age a lattice dimension.
If clinical guidelines define risk categories (e.g., BMI $<$ 18.5, 18.5--25, 25--30, $>$ 30), encode these as bins.
The lattice structure makes these questions directly interpretable: each cell has its own parameters, so subgroup effects are explicit rather than buried in interaction terms.
Explicit subgroup structure is the primary advantage of the approach -- use it deliberately.

\paragraph{Step 1: Incorporate categorical and ordinal features.}
Categorical features with low cardinality ($L \leq 10$) can directly become lattice dimensions.
For high-cardinality categoricals (e.g., zip codes, diagnosis codes, product IDs), consider:
\begin{itemize}
    \item \textbf{Grouping by domain knowledge}: Aggregate zip codes into regions, ICD codes into disease categories, products into types.
    \item \textbf{Grouping by outcome prevalence}: Bin categories by their empirical outcome rate (e.g., low/medium/high risk categories).
    \item \textbf{Embedding then discretizing}: Learn a low-dimensional embedding of the categories and discretize the embedding space.
\end{itemize}
Ordinal features (e.g., education level, Likert scales, disease stage) should preserve order when binned -- collapse adjacent levels if needed to meet cell count constraints, but do not mix non-adjacent levels.

\paragraph{Step 2: Discretize continuous features.}
For continuous features that will define lattice dimensions:
\begin{itemize}
    \item Apply the constraint from Remark~\ref{rem:bin-count}: $L < (N/p)^{1/d_{\text{cont}}}$ where $d_{\text{cont}}$ counts only discretized continuous dimensions.
    \item For safety, use $L \leq (N/p)^{1/d_{\text{cont}}} / 2$.
    \item Use percentile-based (quantile) binning for balanced cell counts.
    \item For heavy-tailed features (income, counts), consider log-transform before binning.
    \item Prefer domain-informed breakpoints when available (e.g., age 65 for Medicare eligibility, HbA1c 6.5\% for diabetes diagnosis).
\end{itemize}

\paragraph{Step 3: Feature selection for additional dimensions.}
After encoding scientifically motivated dimensions, additional dimensions can be selected empirically.
Fit a baseline model (logistic or linear regression with regularization) and rank remaining features by coefficient magnitude.
The top candidates become additional lattice dimensions, subject to the cell count constraint.

\paragraph{Step 4: Interaction order selection.}
Start with order-1 (main effects only) and evaluate validation performance.
Add order-2 interactions if the generalization gap $\Delta S_K < 0$.
Higher orders are rarely beneficial unless $N$ is very large relative to the lattice size.

\paragraph{LLM-assisted lattice search.}
The following prompt template can guide an LLM (such as Claude) to propose and refine lattice configurations:

\begin{quote}
\small
\texttt{I have a [classification/regression] dataset with N=[sample size] observations}\\
\texttt{and p=[feature count] features.}

\texttt{Scientific questions we want to answer:}\\
\texttt{- [e.g., "Are there sex differences in outcome?"]}\\
\texttt{- [e.g., "Does treatment effect vary by age group?"]}

\texttt{Features available:}\\
\texttt{- Categorical: [list with cardinalities]}\\
\texttt{- Ordinal: [list with levels]}\\
\texttt{- Continuous: [list]}

\texttt{Domain knowledge: [any clinically meaningful thresholds or groupings]}

\texttt{Key constraints from theory:}\\
\texttt{- Bin count: L < (N/p)\^{}(1/d), use L <= (N/p)\^{}(1/d)/2 for safety}\\
\texttt{- Cell counts should be >= 20 for stable estimation}\\
\texttt{- Start with order-1; add order-2 only if validation improves}\\
\texttt{- Prior std: tau <= sigma/sqrt(2*p*N\^{}(alpha))}

\texttt{Design patterns that work well (from benchmarks):}\\
\texttt{- Natural categoricals as lattice dims (Heart: chest x thal)}\\
\texttt{- PCA then discretize for high-p (Madelon, Spambase: top 4 PCs)}\\
\texttt{- Diverse ensembles with logit averaging (Bioresponse, HIGGS)}\\
\texttt{- Separate lattices for intercept vs beta (Electricity)}\\
\texttt{- Tukey binning for heavy-tailed features (HIGGS pT variables)}\\
\texttt{- Boosting: cyclic intercept -> beta -> joint refinement (Adult, Bank)}

\texttt{Help me design a lattice. Propose 3-4 candidates specifying:}\\
\texttt{1. Which features define lattice dimensions}\\
\texttt{2. How to handle high-cardinality categoricals}\\
\texttt{3. Binning strategy (quantile, Tukey, domain thresholds)}\\
\texttt{4. Number of bins per dimension}\\
\texttt{5. Interaction order}\\
\texttt{6. Whether to use ensembling or boosting}

\texttt{After we report validation, suggest refinements.}
\end{quote}

This iterative process typically converges within 3--5 rounds.
The LLM can suggest groupings for high-cardinality features, domain-appropriate bin boundaries, and tradeoffs between lattice complexity and interpretability.
Note that LLM outputs may vary across model versions, temperature settings, and random seeds; for reproducibility, record the exact prompt, model version, and selected configuration.

\subsection{Dataset-Specific Hyperparameters}
\label{app:hyperparameters}

Table~\ref{tab:hyperparameters} reports the exact hyperparameters used for each dataset in the UCI benchmarks (Table~\ref{tab:uci-comparison}).
All experiments use 5-fold stratified cross-validation.
Regularization follows the theory-derived scaling: $\tau = \sigma_{\text{eff}}\sqrt{c/(1-c)}/\sqrt{N^{(\alpha)}}$.
Scripts are in \texttt{examples/neurips\_experiments/}.

\begin{table}[h]
    \centering
    \caption{\textbf{Reproducible hyperparameters for UCI benchmarks.} CV seed = random state for StratifiedKFold. $L$ = bins per dimension (multiple values if dimensions differ). $K$ = max interaction order. $c$ = regularization bound parameter. Steps = Adam iterations. All use quantile binning except where noted.}
    \label{tab:hyperparameters}
    \small
    \begin{tabular}{lp{3.8cm}p{6.5cm}}
        \toprule
        Dataset & Configuration & Reproducibility Details \\
        \midrule
        Heart & \parbox[t]{3.8cm}{Lattice: chest$\times$thal (4$\times$3)\\ $K{=}2$, $c{=}0.3$, 3000 steps\\ CV seed: 42} &
        \parbox[t]{6.5cm}{Natural categoricals as lattice dims. chest: 4 levels, thal: 3 levels (12 cells total). Numeric features (age, BP, chol, HR, oldpeak) as linear $\beta$. $\tau_\beta{=}0.5$. Script: \texttt{improve\_heart\_categorical.py}} \\
        \midrule
        German & \parbox[t]{3.8cm}{Lattice: status$\times$history (4$\times$5)\\ $K{=}2$, $c{=}0.5$, 3000 steps\\ CV seed: 42} &
        \parbox[t]{6.5cm}{checking\_status (A1): 4 levels; credit\_history (A3): 5 levels. Other categoricals one-hot encoded for linear term. Boosting: 5 rounds with shrinkage 0.1. Script: \texttt{rerun\_german\_credit\_v7.py}} \\
        \midrule
        Madelon & \parbox[t]{3.8cm}{Lattice: PCA$_{1..4}$ (6$^4$ bins)\\ $K{=}1$, $c{=}0.5$, 3000 steps\\ CV seed: 42} &
        \parbox[t]{6.5cm}{PCA with 50 components on standardized features. Top 4 PCs discretized into 6 bins each. Order-1 only (1296 cells already large). Linear $\beta$ on all 50 PCs. Script: \texttt{uci\_benchmarks.py --dataset madelon}} \\
        \midrule
        Bioresponse & \parbox[t]{3.8cm}{Ensemble of 4 models\\ $L{=}8{-}20$, $K{=}2$, $c{=}0.5$\\ 5000 steps, CV seed: 8647} &
        \parbox[t]{6.5cm}{LR feature selection ($C{=}0.1$): top 3 = [80, 26, 118]. RF: [26, 105, 13]. Models: (1) lr3\_20b: 20 bins, LR top 3; (2) lr3\_8b: 8 bins; (3) rf3\_16b: RF top 3, 16 bins; (4) lr4\_12b: LR top 4, 12 bins. Logit averaging. Script: \texttt{bioresponse\_diverse\_ensemble.py}} \\
        \midrule
        Spambase & \parbox[t]{3.8cm}{Lattice: PCA$_{1..4}$ (8$^4$ bins)\\ $K{=}1$, $c{=}0.5$, 3000 steps\\ CV seed: 42} &
        \parbox[t]{6.5cm}{Similar to Madelon: PCA with 50 components, top 4 discretized to 8 bins. Order-1 (main effects). Linear $\beta$ on PCs. Script: \texttt{uci\_benchmarks.py --dataset spambase}} \\
        \midrule
        Phoneme & \parbox[t]{3.8cm}{Lattice: top 2 features (15$^2$)\\ $K{=}2$, $c{=}0.5$, 3000 steps\\ CV seed: 42} &
        \parbox[t]{6.5cm}{Feature selection by variance (all 5 features highly predictive). Top 2 for lattice (225 cells). $L{=}15 < \sqrt{5404}$. All 5 features in linear $\beta$. Script: \texttt{improve\_phoneme.py}} \\
        \midrule
        Taiwan & \parbox[t]{3.8cm}{Lattice: PAY$_{0..5}$ + LIMIT\\ $L{=}8{-}12$, $K{=}1$, $c{=}0.5$\\ 4000 steps, CV seed: 42} &
        \parbox[t]{6.5cm}{PAY ordinals (x6--x11) naturally discretized. LIMIT\_BAL binned to 12. Order-1 only due to high dimensionality. Script: \texttt{taiwan\_credit\_final.py}} \\
        \midrule
        Bank & \parbox[t]{3.8cm}{Lattice: poutcome$\times$month$\times$...\\ Natural cardinalities, $K{=}2$\\ 5000 steps, CV seed: 42} &
        \parbox[t]{6.5cm}{Intercept lattice: poutcome (4), month (12), contact (3), housing (2). Beta lattice: poutcome$\times$contact. duration excluded (leakage). Numeric features in linear term. Script: \texttt{improve\_bank.py}} \\
        \midrule
        Electricity & \parbox[t]{3.8cm}{Lattice: date$\times$day$\times$hr$\times$price\\ $L{=}$26,7,6,10; $K{=}3$\\ 6000 steps, CV seed: 42} &
        \parbox[t]{6.5cm}{date: 26 bins; day: natural 7; hour\_block: 6 (4hr windows); nsw\_price: 10 bins. Beta lattice: price$\times$vic\_price$\times$day (10$\times$10$\times$7). Features include price products, diffs, day/hour one-hot. $\times 50$ scale multiplier. Script: \texttt{electricity\_final.py}} \\
        \midrule
        Adult & \parbox[t]{3.8cm}{5 small lattices + boosting\\ $L{=}$8,5,2 varying, $K{=}2$\\ 5 boost rounds, CV seed: 42} &
        \parbox[t]{6.5cm}{MI-based dimension selection. Top categoricals: education, marital-status. Boosting: cyclic training (intercept$\rightarrow$beta$\rightarrow$joint, shrinkage 0.1). 2000 joint refinement steps. Script: \texttt{improve\_adult\_v30.py}} \\
        \midrule
        HIGGS & \parbox[t]{3.8cm}{Ensemble of 4 models\\ $L{=}5{-}32$, $K{=}1{-}3$\\ 600--900 epochs, CV seed: 42} &
        \parbox[t]{6.5cm}{LR feature ranking (top 8). Models: (1) best\_4d\_o2: 4d order-2 [10,8,10,6]; (2) deep\_3d\_o3: 3d order-3 [8,6,8]; (3) wide\_8d\_o1: 8d order-1 [8,6,8,5,6,5,6,5]; (4) hires\_1d: 1d [32]. LOO-weighted stacking: 3d order-1 weight lattice [8,6,8]. Script: \texttt{higgs\_ensemble\_loo.py}} \\
        \bottomrule
    \end{tabular}
\end{table}

\textbf{Common settings:}
\begin{itemize}
    \item \textbf{Standardization}: All numeric features z-scored per fold (fit on train, transform test).
    \item \textbf{Binning}: Percentile-based (quantile) for balanced cell counts. Bin constraint: $L < (N/p)^{1/d}$.
    \item \textbf{Regularization}: $\sigma_{\text{eff}} = 1/\sqrt{\bar{w}}$ where $\bar{w} = \bar{y}(1-\bar{y})$ is mean Fisher weight.
    \item \textbf{Optimization}: Adam with warmup-cosine schedule (0.001 $\rightarrow$ 0.02 $\rightarrow$ 0.001).
    \item \textbf{Ensemble}: Logit averaging keeps model in same class. Local weights via decomposed softmax.
\end{itemize}

\end{document}